\documentclass{IEEEtran}
\usepackage{amssymb}
\usepackage{amsfonts}
\usepackage{amsmath}
\usepackage{algorithm}
\usepackage{graphicx,subfigure,cite,multicol,multirow,diagbox,booktabs,array}
\usepackage{color}

\ifCLASSINFOpdf
\else
\fi

\begin{document}
%
\title{ Performance Analysis of Directional Modulation with Finite-quantized RF Phase Shifters in Analog Beamforming Structure \\ }

\author{Jiayu Li,~
        Ling~Xu,~Ping~Lu,~Tingting~Liu,~\IEEEmembership{Member,~IEEE},
        ~Zhihong~Zhuang,~Jinsong~Hu,~Feng~Shu~and~Jiangzhou Wang,~\IEEEmembership{Fellow,~IEEE}

\thanks{This work was supported in part by the National Natural Science Foundation of China (Nos. 61771244, 61472190, 61501238, and 61702258).}
\thanks{Jiayu Li,~
        Ling~Xu,~Zhihong~Zhuang~and~Feng Shu are with the School of Electronic and Optical Engineering, Nanjing University of Science and Technology, 210094, CHINA.}
\thanks{Ping Lu is with Zhongxing Telecommunication Equipment Corporation. Email: lu.ping@zte.com.cn.}
\thanks{Tingting~Liu is with the School of information and communication engineering, Nanjing Institute of Technology, Nanjing 211167, China. Email: liutt@njit.edu.cn.}
\thanks{Jinsong~Hu is with the College of Physics and Information, Fuzhou University, Fuzhou 350116, ~China.}
\thanks{Jiangzhou Wang is with the School of Engineering and Digital Arts, University of Kent, Canterbury CT2 7NT, U.K. Email: \{j.z.wang\}@kent.ac.uk.}

}
\maketitle

\begin{abstract}
The radio frequency (RF) phase shifter with finite quantization bits in analog beamforming (AB) structure forms quantization error (QE) and causes a performance loss of received signal to interference plus noise ratio (SINR) at the receiver (called Bob). By using the law of large numbers in probability theory, the closed-form expression of SINR performance loss is derived to be inversely proportional to the square of sinc (or $\sin(x)/x$) function. Here, a phase alignment method is applied in directional modulation transmitter with AB structure. Also, the secrecy rate (SR) expression is derived with QE. From numerical simulation results, we find that the SINR performance loss gradually decreases as the number $L$ of quantization bits increases. This loss is less than 0.3dB when $L$ is larger than or equal to 3. As $L$ exceeds 5, the SINR performance loss at Bob can be approximately trivial. Similarly, SR performance loss gradually reduces as $L$ increases. In particular, the SR performance loss is about 0.1 bits/s/Hz for $L=3$ at signal-to-noise ratio of $15$dB.
\end{abstract}

\maketitle

\section{Introduction}
\IEEEPARstart{D}{irectional} modulation (DM), as one of the key technologies of wireless physical layer security, is attracting ever-increasing research interests and activities from both academia and industry world.
Traditional technology for directional modulation was proposed on the radio frequency (RF) frontend \cite{DM-PA,DM-Array,DM-securearray}. In these articles, the authors proposed an actively driven DM array of utilizing analog RF phase shifters or antenna elements, which did not deal with the flexibility of design process.
Another way to implement the DM synthesis is based on the baseband signal processing. In \cite{YD-avector}, the authors proposed to form an orthogonal vector, which can be updated in the null space of channel vector at the desired direction, to the transmitted baseband signal as artificial noise (AN), thereby improving the secure transmission. Compared to the design on the RF frontend, this approach enables dynamic DM transmissions and makes the design easier.

In the presence of direction measurement error, the authors in \cite{HU2016-RDM}, \cite{WU2016-RDMBC} and \cite{MU-MIMO-ZHU} proposed three robust DM synthesis methods for three different scenarios: single-desired user, multi-user broadcasting and multi-user multi-input multi-output (MIMO) by fully exploiting the statistical properties of direction measurement error.
\cite{DM-XU} proposed two secure schemes, Max-GRP plus NSP and Max-SLNR plus Max-ANLNR, for multicast DM scenario to improve the security. Inspired by the work in \cite{WANG-OFDMA1} and \cite{WANG-OFDMA2}, secure and precise wireless transmission (SPWT) proposed in \cite{Wu-SPTDM} combined AN projection, beamforming and random subcarrier selection based on orthogonal frequency division multiplexing (OFDM) to achieve SPWT of confidential messages. In the researches mentioned above, the DM synthesis on the baseband signal processing is assumed perfect or imperfect channel state information (CSI). In \cite{DOA-QIN}, the authors proposed three estimators of directions of arrival (DOA) based on hybrid structure for finding direction, thereby determining the position. This method makes DM more practical.

In \cite{HU2016-RDM}, \cite{WU2016-RDMBC}, and \cite{MU-MIMO-ZHU}, the authors proposed robust methods for imperfect CSI in traditional DM systems, i.e, fully-digital (FD) beamforming systems. Traditional fully-digital beamforming technique is of high cost and power consumption due to each antenna element requiring one dedicated RF chain. Hybrid analog/digital (HAD) beamforming structure \cite{O.E.HP-SSP,A.O.HP,Y.C-HP} with analog phase shifters and a reduced number of RF chains was proposed to strike a  good balance between the system complexity and the beamforming precision. Compared to HAD and FD beamforming structures, analog beamforming (AB) structure with digitally-controlled phase shifters has attracted substantial research attentions from both industry and academic communities, due to its low circuit cost and high energy efficiency \cite{beamcodebook,anabeam-phaseshift,cdtbeam,S-ANB-base}. In general, AB structure has only single RF chain linked to all antennas. However, AB as described in \cite{anabeam-phaseshift,S-ANB-base} is subject to additional constraints, for example, the digitally-controlled phase shifters with finite-quantized phase values and constant-envelope. Here, due to finite-quantized phase values, there exists quantization error (QE), which will lead to a performance loss such as signal to interference plus noise ratio (SINR) and secrecy rate (SR). It is crucial to derive and analyze the impact of QE on SINR and SR due to the accuracy of quantization of phase shifter. To achieve an allowable performance loss, what is the minimum number of quantization bits compared with infinite-bit quantization (no QE, NQE)? In what follows, we will address this issue.

In this paper, we will mainly present analysis of the effect of QE from finite-quantized phase shifters on the performance of DM system using AB structure. Here, the transmitter Alice is equipped with an AB structure, while the desired receiver at Bob works in full-duplex model and helps Alice by transmitting AN with FD  beamforming structure to degrade the performance of the illegitimate receiver at Eve. The main contributions of this paper are summarized as follows:
\begin{enumerate}
 \item  In AB structure, the RF phase shifter usually has finite quantization bits. This will result in a receive SINR performance loss at Bob. By using the law of large numbers in probability theory, the approximate closed-form expression of SINR performance loss is derived to be inversely proportional to the square of sinc (called $\sin(x)/x$) function. This will greatly simplify the analysis that how many bits is sufficient such that the SINR performance loss can be omitted in the AB structure.

  \item  From simulation results, it follows that this approximate expression holds even for a small-scale number of transmit antennas at Alice.  Additionally, we also find an important result that the SINR performance loss is less than 0.3dB when the number $L$ of quantization bits is larger than or equal to 3. As the number of quantization bits exceeds 4, the SINR performance loss at Bob can be completely negligible.

 \item  In the presence of QE, the expression of SR is also derived and simplified. Simulation results indicate that the SR performance loss is about 0.1 bits/s/Hz when $L=3$. More importantly, as the value of $L$ increases, the SR performance loss decreases gradually and monotonically.  Thus, $L=3$ is sufficient for RF phase quantizer in the AB structure.
\end{enumerate}

The remainder of this paper is organized as follows. Section \ref{S2} describes the system model. In Section \ref{S3}, the expression of SINR loss is derived by modeling quantization error as a uniform distribution, and at the same time the corresponding SR expression is given in the presence of QE. Simulation results are presented in Section \ref{S4}. Finally, we make our conclusions in Section \ref{S5}.

\emph{Notations:} throughout the paper, matrices, vectors, and scalars are denoted by letters of bold upper case, bold lower case, and lower case, respectively. Signs $(\cdot)^T$, $(\cdot)^H$ and $\mid\cdot\mid$ denote transpose, conjugate transpose, and modulus respectively. Notation $\mathbb{E}\{\cdot\}$ stands for the expectation operation.

\section{System Model}\label{S2}
\begin{figure*}[htb]
  \centering
  \includegraphics[width=0.8\textwidth]{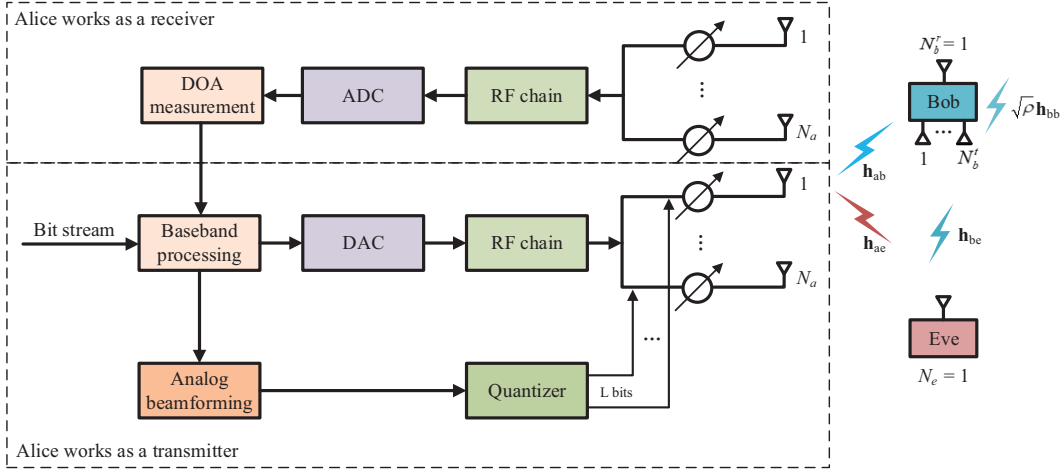}\\
  \caption{System model.}
  \label{analogmodel}
\end{figure*}

Consider a DM network with a Gaussian wiretap channel in Fig.~\ref{analogmodel}, where Alice is equipped with $N_a$ antennas, Bob is equipped with $N_b$ antennas, and Eve is equipped with single antenna. Alice intends to send its confidential message $\mathbf{x}$ to Bob, without being wiretapped by Eve. The DM transmitter at Alice adopts an AB structure. This means Alice can  send single confidential message stream to Bob by analog beamforming due to only one RF chain. In order to help Alice, Bob operates in a FD mode. In other words,  all antennas at Bob are partitioned into two subsets. The first subset of antennas with $N_b^t$ antennas transmits AN $\mathbf{z}$, and the second one with $N_b^r\ =N_b\ - \ N_b^t\ $ antennas receives confidential messages from Alice. It is supposed $ N_b^t\ =1 $ so that Bob owns single antenna to receive as Eve. Since Bob transmits AN while receiving the desired signal, there always exists self-interference at its own receive signal. To describe the effect of residual self-interference we employ the loop interference model of \cite{FD-REC-GZ}, which quantifies the level of self-interference with a parameter $\rho \in [0,1]$, with $\rho = 0$ denoting zero self-interference. In this paper, we assume there exists the line-of-sight (LOS) path. The transmit signal at Alice and AN at Bob can be respectively written as
\begin{align}\label{sa signal}
\mathbf{s}_a=\sqrt{P_a}\mathbf{v}_ax,
\end{align}
and
\begin{align}\label{sb signal}
\mathbf{s}_b=\sqrt{P_b}\mathbf{v}_bz,
\end{align}
where $P_a$ and $P_b$ are the transmission powers of Alice and Bob, respectively. Vector
\begin{align}\label{va}
\mathbf{v}_a(\alpha)=\frac{1}{\sqrt{N_a}}\left[e^{j\hat\alpha_1},e^{j\hat\alpha_2},\cdots,e^{j\hat\alpha_{N_a}}\right]^{T}
\end{align}
denotes the transmit analog beamforming vector, which forces the confidential message to the desired direction and $\mathbf{v}_b\in\mathbb{C}^{N_b\times1}$ is the beamforming vector of  transmitting AN to interfere with Eve. An AB pattern is generated by a digitally-controlled RF phase-shifter with $L$-bit phase quantizer. This means that each antenna's phase in (\ref{va}) takes one nearest value $\hat\alpha_n$ to the designed value $\alpha_n$  from a set of  $2^L$ quantized phases given by
\begin{align}\label{alphan}
\hat\alpha_n\in \Theta=\bigg\{ 0, 2\pi(\frac{1}{2^L}), 2\pi(\frac{2}{2^L}),\cdots, 2\pi(\frac{2^L-1}{2^L}) \bigg\},
\end{align}
which is actually an integer optimization problem. Therefore, the beamforming vector in the AB system is defined with the quantized phases $\alpha_n$ and written as (\ref{va}). Each element phase is quantized to $L$ bits.
In (\ref{sa signal}),  $x$ is the confidential message of satisfying $\mathbb{E}\left\{x^Hx\right\}=1$ . We assume that the AN $z$ transmitted by Bob obeys a Gaussian distribution with zero mean and $\mathbb{E}\left\{z^Hz\right\}=1$.

Taking the path loss into consideration, the signal received at Bob and Eve can be respectively written as
\begin{align}\label{yb}
y_b
&=\sqrt{g_{ab}}\mathbf{h}_{ab}^{H}(\theta_d)\mathbf{s}_a+\sqrt\rho\mathbf{h}_{bb}^{H}\mathbf{s}_b+n_b\\\nonumber
&=\sqrt{g_{ab}P_a}\mathbf{h}_{ab}^{H}(\theta_d)\mathbf{v}_ax+\sqrt{\rho P_b}\mathbf{h}_{bb}^{H}\mathbf{v}_bz+n_b,
\end{align}
and
\begin{align}\label{ye}
y_e
&=\sqrt{g_{ae}}\mathbf{h}_{ae}^{H}(\theta_e)\mathbf{s}_a+\sqrt{g_{be}}\mathbf{h}_{be}^{H}\mathbf{s}_b+n_e\\\nonumber
&=\sqrt{g_{ae}P_a}\mathbf{h}_{ae}^{H}(\theta_e)\mathbf{v}_ax+\sqrt{g_{be}P_b}\mathbf{h}_{be}^{H}\mathbf{v}_bz+n_e,
\end{align}
where $g_{ab}=\frac{\epsilon}{d_{ab}^c}$ and $d_{ab}$ denote the loss coefficient and distance between Alice and Bob respectively. $c$ is the path loss exponent and $\epsilon$ is the attenuation at reference distance $d_0$. Likewise, $g_{ae}=\frac{\epsilon}{d_{ae}^c}$ and $d_{ae}$ denote the loss coefficient and distance between Alice and Eve, respectively. $g_{be}=\frac{\epsilon}{d_{be}^c}$ and $d_{be}$ denote the loss coefficient and distance between Bob and Eve, respectively.
$n_b\sim\mathcal{C}\mathcal{N}(0,\sigma_b^2)$ and $n_e\sim\mathcal{C}\mathcal{N}(0,\sigma_e^2)$ represent complex additive white Gaussian noise (AWGN) at Bob and Eve, respectively. $\mathbf{h}_{ab}\in\mathbb{C}^{N_a\times1}$ denotes the channel vector from Alice to Bob, $\mathbf{h}_{ae}\in\mathbb{C}^{N_a\times1}$ and $\mathbf{h}_{be}\in\mathbb{C}^{N_b^t\times1}$ denote the channel vectors from Alice and Bob to Eve, respectively. $\mathbf{h}_{bb}\in\mathbb{C}^{N_b^t\times1}$ represents the self-interference channel vector at Bob. In the following, we assume that $\sigma_b^2=\sigma_e^2=\sigma^2$ .

In Fig.~1, the transmitter is deployed with an $N_a$-element linear antenna array. The normalized steering vector (NSV) for the transmit antenna array is denoted by
\begin{align}\label{h_theta}
\mathbf{h}(\theta)=\left[e^{j2\pi\Psi_{\theta}(1)}, \cdots, e^{j2\pi\Psi_{\theta}(n)}, \cdots, e^{j2\pi\Psi_{\theta}(N_a)}\right]^T,
\end{align}
and the phase function $\Psi_{\theta}(n)$ is defined as
\begin{align}\label{var_phi}
\Psi_{\theta}(n)\triangleq-\frac{(n-(N_a+1)/2)d\cos\theta}{\lambda}, n=1,2,\cdots,N_a,
\end{align}
where $\theta$ is the direction angle, $n$ denotes the $n$-th antenna, $d$ is the distance of two adjacent antennas, and $\lambda$ is the wavelength. Making use of the definition of NSV, we have $\mathbf{h}_{ab}(\theta_d)=\mathbf{h}(\theta_d)$ and $\mathbf{h}_{ae}(\theta_e)=\mathbf{h}(\theta_e)$.

If the beamforming vector $\mathbf{v}_a$ is determined, the optimal $\mathbf{v}_b$ can be solved by using the Max-SR method \cite{GPI-HY} and utilizing the GPI algorithm\cite{GPI-N}.

\section{Derivation of SINR and SR performance loss expressions}\label{S3}
In this paper, we focus on the impact of quantization error of the phase shifter on SINR and SR performance, which will cause phase mismatch between the NSV $\mathbf{h}$ and the AB vector even with ideal measurement of direction. This will degrade the receive performance at Bob, including the receive SINR loss and SR reduction. The small QE in the phase shifter may severely degrade the performance of the DM system. To analyze this problem, in this section, we assume that $\theta_d$ is randomly chosen from the interval $ [0,360^o)$.  Let us denote $\alpha_n$ by the designed or ideal AB phase of antenna $n$ at Alice. Considering the  effect of QE, we establish the model of QE as follows
\begin{align}\label{alphahat}
\widehat{\alpha}_n=\alpha_n+\Delta\alpha_n,~n\in 1,2,\cdots,N_a,
\end{align}
where $\widehat{\alpha}_n\in\Theta$ is the quantized value of $\alpha_n$ after $\alpha_n$ passes through the corresponding phase quantizer. In the above model, the quantization error $\Delta\alpha_n$ is approximated as a uniform distribution and its probability density function (PDF) is given by
\begin{align}\label{pDeltaalpha}
p(\Delta\alpha_n)=
\begin{cases}
\frac{1}{2\Delta\alpha_{max}},  &\Delta\alpha_n\in\left[-\Delta\alpha_{max},~\Delta\alpha_{max}\right],\\
0,  & \mbox{otherwise},
\end{cases}
\end{align}
with
\begin{align}\label{deltaalpha}
\Delta\alpha_{max}=\frac{\pi}{2^L},
\end{align}
where $L$ is the number of quantization bits.
\subsection{Derivation of SINR Loss due to finite-bit quantization}
Given the predesigned AB vector $\mathbf{v}_a(\alpha)$, we have
\begin{align}\label{vahat}
\mathbf{v}_a(\widehat{\alpha})
&=\frac{1}{\sqrt{N_a}}\left[e^{j\widehat{\alpha}_1},e^{j\widehat{\alpha}_2},\cdots,e^{j\widehat{\alpha}_{N_a}}\right]^{T}\\\nonumber
&=\frac{1}{\sqrt{N_a}}\left[e^{j(\alpha_1+\Delta\alpha_1)},e^{j(\alpha_2+\Delta\alpha_2)},\cdots,e^{j(\alpha_{N_a}+\Delta\alpha_{N_a})}\right]^{T}.
\end{align}
Substituting the above in (\ref{sa signal}), the RF transmit signal at Alice can be rewritten as
\begin{align}\label{sanon}
\mathbf{s}_a(\widehat{\alpha})=\sqrt{P_a}\mathbf{v}_a(\widehat{\alpha})x.
\end{align}
In this case, the corresponding receive signals at Bob and Eve can be respectively written as
\begin{align}\label{ybnon}
y_b(\widehat{\alpha})
&=\sqrt{g_{ab}}\mathbf{h}_{ab}^{H}(\theta_{d})\mathbf{s}_a(\widehat{\alpha})+\sqrt\rho\mathbf{h}_{bb}^{H}\mathbf{s}_b+n_b\\\nonumber
&=\sqrt{g_{ab}P_a}\mathbf{h}_{ab}^{H}(\theta_{d})\mathbf{v}_a(\widehat{\alpha})x+\sqrt{\rho P_b}\mathbf{h}_{bb}^{H}\mathbf{v}_bz+n_b,
\end{align}
and
\begin{align}\label{yenon}
y_e(\widehat{\alpha})
&=\sqrt{g_{ae}}\mathbf{h}_{ae}^{H}(\theta_{e})\mathbf{s}_a(\widehat{\alpha})+\sqrt{g_{be}}\mathbf{h}_{be}^{H}\mathbf{s}_b+n_e\\\nonumber
&=\sqrt{g_{ae}P_a}\mathbf{h}_{ae}^{H}(\theta_{e})\mathbf{v}_a(\widehat{\alpha})x+\sqrt{g_{be}P_b}\mathbf{h}_{be}^{H}\mathbf{v}_bz+n_e.
\end{align}
Assuming that the ideal desired directional angle $\theta_d$ is available, we have
\begin{align}\label{alphaand}
\alpha_n=2\pi\Psi_{\theta_d}(n),~\hat\alpha_n=2\pi\Psi_{\theta_d}(n)+\Delta\alpha_n.
\end{align}
Substituting the above two equations in (\ref{ybnon}) and (\ref{yenon}) yields
\begin{align}\label{habHv}
\mathbf{h}_{ab}^{H}(\theta_{d})\mathbf{v}_a(\widehat{\alpha})
&=\big[e^{-j\alpha_{1}},e^{-j\alpha_{2}},\cdots,e^{-j\alpha_{N_a}}\big]\\\nonumber
&\times\frac{1}{\sqrt{N_a}}\big[e^{j(\alpha_{1}+\Delta\alpha_{1})},e^{j(\alpha_{2}+\Delta\alpha_{2})},\\\nonumber
& \cdots, e^{j(\alpha_{N_a}+\Delta\alpha_{N_a})}\big]^{T}\\\nonumber
&=\frac{1}{\sqrt{N_a}}\sum_{n=1}^{N_a} e^{j\Delta\alpha_{n}},
\end{align}
and
\begin{align}\label{haeHv}
\mathbf{h}_{ae}^{H}(\theta_e)\mathbf{v}_a(\widehat{\alpha})
&=\big[e^{-j\alpha_{ae,1}},e^{-j\alpha_{ae,2}},\cdots,e^{-j\alpha_{ae,N_a}}\big]\\\nonumber
&\times\frac{1}{\sqrt{N_a}}\big[e^{j(\alpha_{1}+\Delta\alpha_{1})},e^{j(\alpha_{2}+\Delta\alpha_{2})},\\\nonumber
& \cdots, e^{j(\alpha_{N_a}+\Delta\alpha_{N_a})}\big]^{T}\\\nonumber
&=\frac{1}{\sqrt{N_a}}\sum_{n=1}^{N_a} e^{j(\alpha_{n}-\alpha_{ae,n}+\Delta\alpha_{n})},
\end{align}
respectively. In (\ref{haeHv}), $\alpha_{n}$ is determined by (\ref{alphaand}), $\alpha_{ae,n}$ can be expressed similarly as (\ref{alphaand}) with known $\theta_{e}$, $\alpha_{ae,n}=2\pi\Psi_{\theta_e}(n)$.

In (\ref{habHv}), $e^{j\Delta\alpha_i}(i=1,2,\cdots,N_a)$ can be viewed as independently identical distributed (iid) random variables, in accordance with the law of large numbers in probability theory. The mean of samples is approximately equal to the mean of the distribution \cite{allofstatistics}. As $N_a$ tends to medium-scale and large-scale, we have
\begin{align}\label{eEe}
\frac{1}{N_a}\sum_{n=1}^{N_a} e^{j\Delta\alpha_{n}} \approx \mathbb{E}(e^{j\Delta\alpha_{n}}),
\end{align}
where
\begin{align}\label{Eesinc}
\mathbb{E}(e^{j\Delta\alpha_{n}})
&=\int_{-\Delta\alpha_{max}}^{\Delta\alpha_{max}} \text{e}^{j\Delta\alpha_n} p(\Delta\alpha_n)\, \text{d}\Delta\alpha_n \\\nonumber
&=\frac{\sin(\Delta\alpha_{max})}{\Delta\alpha_{max}}\\\nonumber
&=\textrm{sinc}(\frac{\pi}{2^L})
\end{align}
with
\begin{align}\label{sincfx}
\text{sinc}(x)=\frac{\sin(x)}{x}.
\end{align}
Combining (\ref{eEe}) and (\ref{Eesinc}), one obtains
\begin{align}\label{eEeapprox}
\frac{1}{N_a}\sum_{n=1}^{N_a} e^{j\Delta\alpha_{n}} \approx \textrm{sinc}(\frac{\pi}{2^L}).
\end{align}

Now, we derive the expression of SINR at Bob under the QE and NQE conditions, respectively. The former has NQE while the latter has QE. From the definition of SINR and (\ref{ybnon}), we have
\begin{align}\label{SNRbideal}
\text{SINR}_b^{NQE}= \frac{g_{ab}{P_a}|\mathbf{h}_{ab}^{H}(\theta_d)\mathbf{v}_a(\alpha)|^2}{\rho{P_b}|\mathbf{h}_{bb}^{H}\mathbf{v}_b|^2 +\sigma^2},
\end{align}
\begin{align}\label{SNRbreal}
\text{SINR}_b^{QE}
&= \frac{g_{ab}{P_a} |\mathbf{h}_{ab}^{H}(\theta_d)\mathbf{v}_a(\widehat{\alpha})|^2}
{\rho{P_b}|\mathbf{h}_{bb}^{H}\mathbf{v}_b |^2 +\sigma^2}\\\nonumber
&=\frac{\mathbb{E}_{\widehat{\alpha}}\big[g_{ab}{P_a} |\mathbf{h}_{ab}^{H}(\theta_d)\mathbf{v}_a(\widehat{\alpha})|^2 \big]}{\rho{P_b}|\mathbf{h}_{bb}^{H}\mathbf{v}_b |^2 +\sigma^2}\\\nonumber
&=\frac{g_{ab}{P_a} \sqrt{N_a}\text{sinc}^2(\frac{\pi}{2^L})}{\rho{P_b}|\mathbf{h}_{bb}^{H}\mathbf{v}_b |^2 +\sigma^2}.
\end{align}
According to (\ref{SNRbideal}) and (\ref{SNRbreal}), let us define the SINR performance loss $\gamma$ as the ratio of $\text{SINR}_b^{NQE}$ to $\text{SINR}_b^{QE}$  at Bob as
\begin{align}\label{gamma}
\boldsymbol{\gamma}
&=\frac{\text{SINR}_b^{NQE}}{\text{SINR}_b^{QE}}\\\nonumber
&=\frac{1}{\text{sinc}^2(\frac{\pi}{2^L})}.
\end{align}
Observing the above expression and considering $L$ is a positive integer, it is clear that increasing the value of $L$, i.e. the number of quantization bits, will reduce the SINR performance loss. In other words, the receive SINR performance will be improved gradually.
\subsection{Expression of SR with finite-bit quantization}
In terms of (\ref{yb}) and (\ref{ye}), the achievable rates at Bob and Eve are as follows
\begin{align}\label{Rb}
R_b= \log_2\left(1+\frac{{g_{ab}P_a} |\mathbf{h}_{ab}^{H}\mathbf{v}_a|^2 }{\rho{P_b}|\mathbf{h}_{bb}^{H}\mathbf{v}_b|^2 +\sigma^2}\right),
\end{align}
and
\begin{align}\label{Re}
R_e= \log_2\left(1+\frac{{g_{ae}P_a} |\mathbf{h}_{ae}^{H}\mathbf{v}_a|^2 }{g_{be}{P_b}|\mathbf{h}_{be}^{H}\mathbf{v}_b|^2 +\sigma^2}\right),
\end{align}
respectively, which yield the following achievable SR
\begin{align}\label{Rs}
R_s
&=\max\left\{0,R_b-R_e\right\}\\\nonumber
&=\max\left\{0,\log_2 \bigg(\frac{MT+g_{ab}P_a T |\mathbf{h}_{ab}^{H}\mathbf{v}_a|^2 }{MT+g_{ae}P_a M |\mathbf{h}_{ae}^{H}\mathbf{v}_a|^2}  \bigg) \right\},
\end{align}
where
\begin{align}\label{MT}
M=\rho{P_b}|\mathbf{h}_{bb}^{H}\mathbf{v}_b|^2 +\sigma^2, \\\nonumber
T=g_{be}{P_b}|\mathbf{h}_{be}^{H}\mathbf{v}_b|^2 +\sigma^2.
\end{align}
In the absence of QE, the corresponding SR is given by
\begin{align}\label{Rsideal}
&R_s^{NQE}\\\nonumber
&=\max\left\{0,R_b^{NQE}-R_e^{NQE}\right\}\\\nonumber
&=\max\left\{0,\log_2 \bigg(\frac{MT+g_{ab}P_a T |\mathbf{h}_{ab}^{H}(\theta_d)\mathbf{v}_a(\alpha)|^2 }{MT+g_{ae}P_a M |\mathbf{h}_{ae}^{H}(\theta_e)\mathbf{v}_a(\alpha)|^2}  \bigg) \right\}.
\end{align}
In the presence of QE, the corresponding SR is presented by
\begin{align}\label{Rsreal}
&R_s^{QE}\\\nonumber
&=\max\left\{0,R_b^{QE}-R_e^{QE}\right\}\\\nonumber
&=\max\left\{0,\log_2 \bigg(\frac{MT+g_{ab}P_a T |\mathbf{h}_{ab}^{H}(\theta_d)\mathbf{v}_a(\widehat{\alpha})|^2 }{MT+g_{ae}P_a M |\mathbf{h}_{ae}^{H}(\theta_e)\mathbf{v}_a(\widehat{\alpha})|^2}  \bigg) \right\}\\\nonumber
&=\max\left\{0,\log_2 \bigg(\frac{MT+g_{ab}P_a T \sqrt{N_a}\text{sinc}^2(\frac{\pi}{2^L}) }{MT+g_{ae}P_a M |\mathbf{h}_{ae}^{H}(\theta_e)\mathbf{v}_a(\widehat{\alpha})|^2}  \bigg) \right\}.
\end{align}

\section{Simulation and Discussion}\label{S4}
In this section, we mainly focus on the evaluation of impact of the number of antennas and quantization bits of phase shifters on performance losses including SINR, SR, and BER in an AB structure. In the simulation, system parameters are chosen as follows: quadrature phase shift keying (QPSK) modulation, the total transmission power $P_a=P_b=70$dBm, the spacing between two adjacent antennas $d=\lambda/2$, $\rho=0.5 $, the distance between Alice and Bob, Alice and Eve, Bob and Eve $d_{ab}=d_{ae}=d_{be}=500m$, the path loss exponent $c=2$, the desired direction $\theta_d=\theta_{ab}=60^{\circ}$, and the eavesdropping direction $\theta_e=\theta_{ae}=120^{\circ}$. The direction angle from Bob to Eve is $\theta_{be}=45^{\circ}$. Alice is equipped with $N_a$ antennas, Bob is equipped with $N_b^t=16$ antennas to transmit AN and $N_b^r=1$ to receive confidential signals from Alice.

\begin{figure}[htb]
  \centering
  \includegraphics[width=0.5\textwidth]{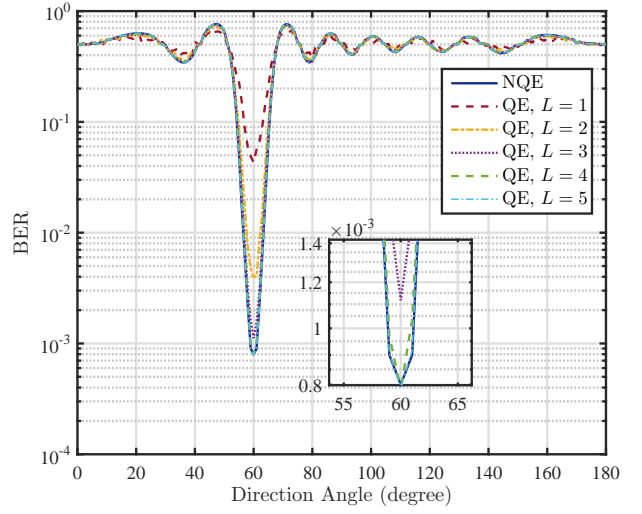}\\
  \caption{Curves of BER versus direction angle under the ideal condition (with NQE) and finite-quantization condition (with QE) for different numbers ($L$) of quantization bits.}
  \label{BER}
\end{figure}

Fig.~\ref{BER} demonstrates the performance curves of bit error rate (BER) versus direction angle at Bob with $\text{SNR}=10\text{dB}$ and $N_a=16$. Here, the ideal condition implies NQE with solid line, i.e., infinite bits for quantization,  and the QE case is denoted by dotted line. $L$ stands for the number of quantization bits. From this figure, it can be seen that the BER can achieve a good performance in the desired direction while it becomes worse rapidly as we move to the undesired direction. This is partly because the AN transmitted from Bob can interfere with the confidential signal received at Eve severely along the undesired directions. Compared with the performance with NQE, the BER performance with QE is much worse, especially for $L\le2$. As $L$ reaches up to 3, the BER performance difference between QE and NQE is trivial. This means that it is feasible in practice to use a finite-quantized phase shifters with $L=3$.

\begin{figure}[htb]
  \centering
  \includegraphics[width=0.5\textwidth]{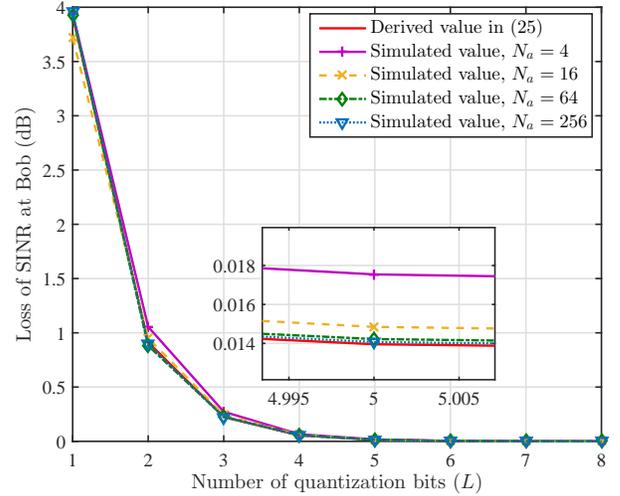}\\
  \caption{SINR performance loss  at Bob versus number $L$ of quantization bits  for different $N_a$.}
  \label{SINRLOSS-BITS}
\end{figure}
\begin{figure}[htb]
  \centering
  \includegraphics[width=0.5\textwidth]{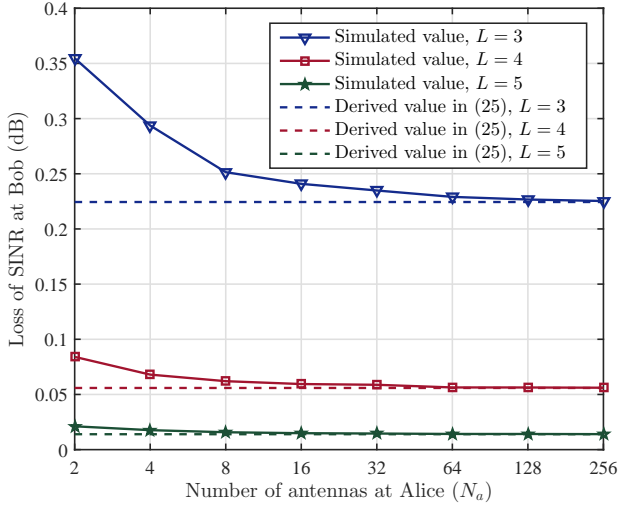}\\
  \caption{SINR performance loss at Bob versus $N_a$ for different numbers of quantization bits ($L$).}
  \label{SINRLOSS-NA}
\end{figure}
\begin{figure}[htb]
  \centering
  \includegraphics[width=0.5\textwidth]{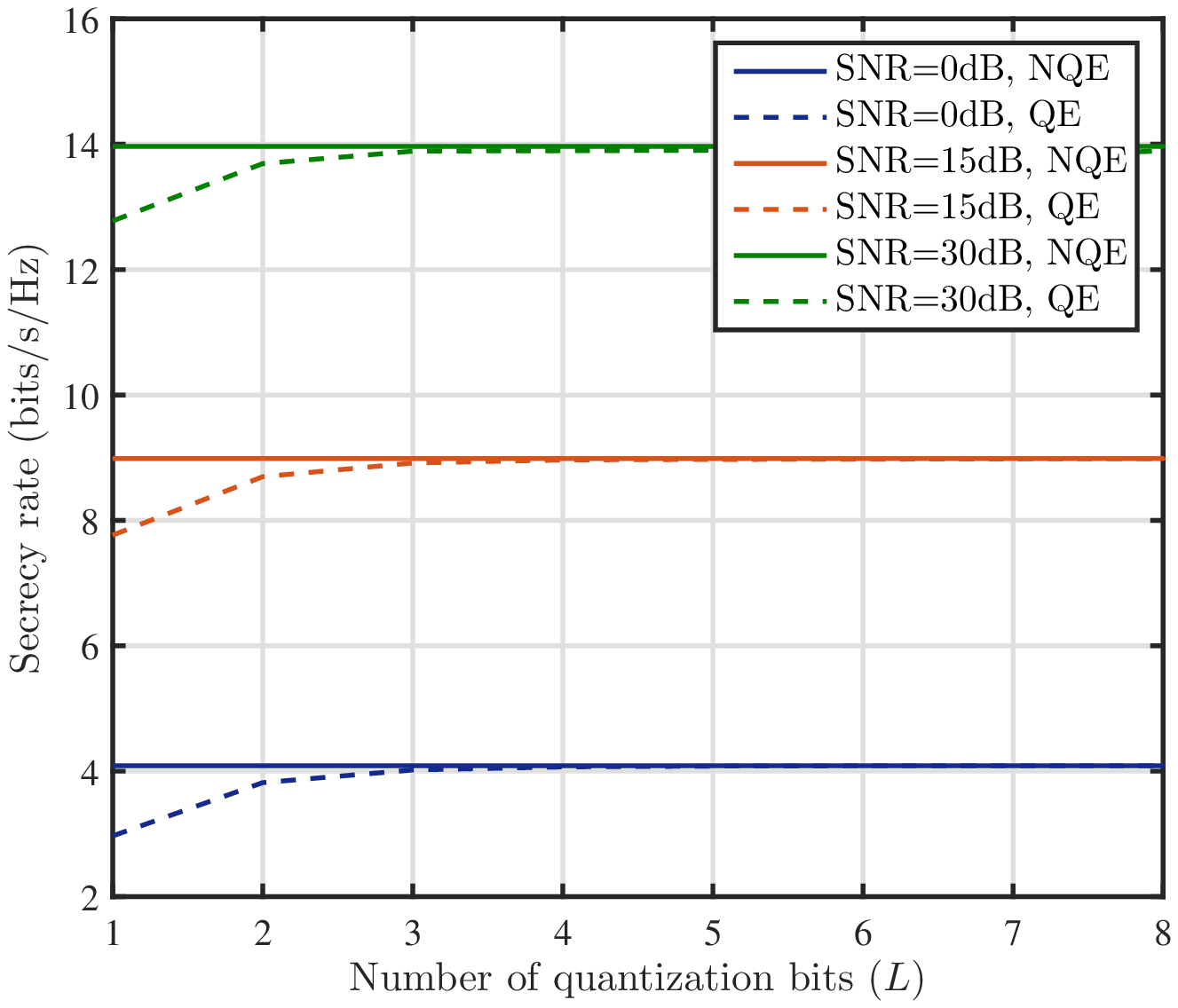}\\
  \caption{Secrecy rate versus number $L$ of quantization bits for different transmit SNR in two cases QE and NQE with $N_a=16$.}
  \label{SR-L-60-120}
\end{figure}

\begin{figure}[!htbp]
  \centering
  \subfigure[{SNR}=0dB]{
  \label {SR-L-Na:SNR:a}
  \includegraphics[width=0.5\textwidth]{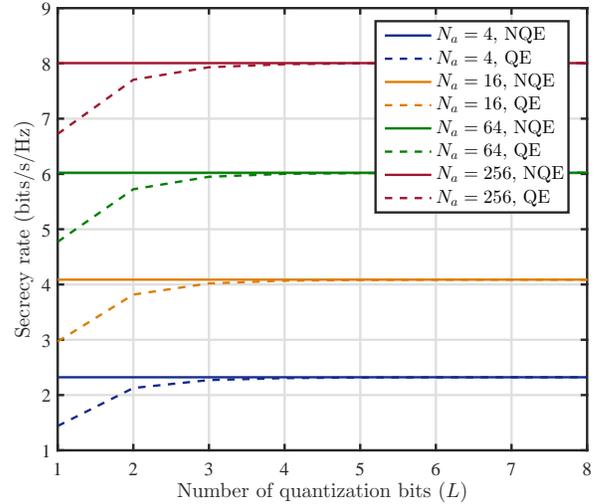}}
  \hspace{0.5in}
  \subfigure[{SNR}=15dB]{
  \label {SR-L-Na:SNR:b}
  \includegraphics[width=0.5\textwidth]{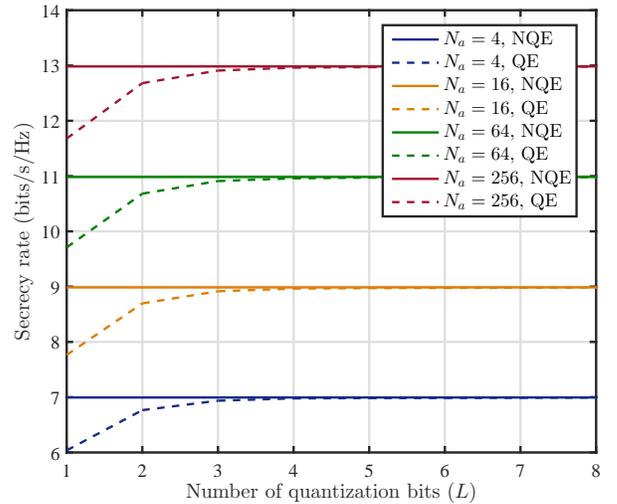}}
  \hspace{0.5in}
  \subfigure[{SNR}=30dB]{
  \label {SR-L-Na:SNR:c}
  \includegraphics[width=0.5\textwidth]{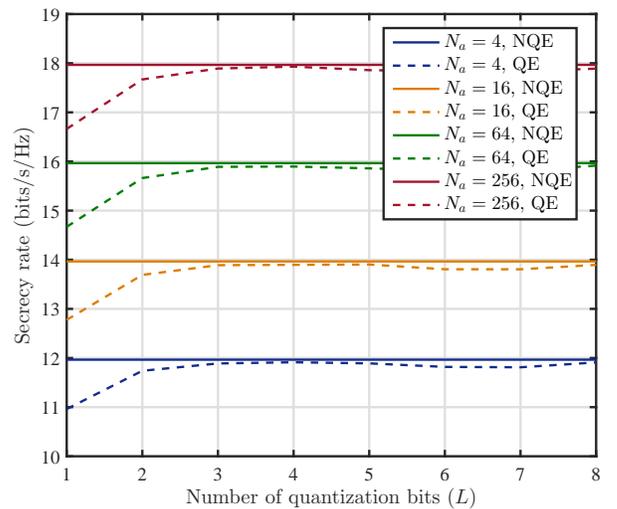}}
  \hspace{0.5in}
  \caption{Secrecy rate versus number $L$ of quantization bits  for different $N_a$ and different transmit {SNR} in two cases {NQE} and {QE}.}
  \label{SR-L-Na:SNR}
\end{figure}

Fig.~\ref{SINRLOSS-BITS} plots the curves of SINR performance loss versus number $L$ of quantization bits  ranging from 1 to 8 for four different numbers of antennas at Alice $N_a$ : 4, 16, 64, and 256, where  SNR is equal to 15dB. Here, the derived expression of SINR performance loss in (\ref{gamma}) is used as a performance reference. From this figure, it is seen that the performance loss of simulated SINR decreases as the quantization bits increases.
This is mainly because that the range of phase error due to quantization (\ref{deltaalpha}) will become smaller as the number  $L$ of quantization bits increases, so that QE will become smaller. This will result in a smaller loss of SINR at Bob. A small number of quantization bits of the phase shifter (e.g., $L=1$ or $2$) will cause a large QE, resulting in a large SINR loss up to 4dB. The SINR performance loss will be less than 0.3dB when the number of quantization bits is more than or equal to 3. When the number of quantization bits is 4, the SINR loss at Bob is less than 0.1dB even if the number of antennas at Alice is small (e.g., $N_a=3$). This also means the fact that even with a small number of antennas at Alice, the derived expression in (\ref{gamma}) coincides with the simulated SINR performance loss. In other words, the derived expression in (\ref{gamma}) can be used to evaluate the SINR performance loss for almost all cases including small-scale, medium-scale, and large-scale. More importantly, we can conclude that three quantization bits are sufficient for the quantized phase shifters in the AB system.

Since we have the approximate derived simple expression for SINR performance loss,  Fig.~\ref{SINRLOSS-NA} illustrates the curves of the SINR performance loss versus the number $N_a$ of antennas at Alice for three different numbers of quantization bits: 3, 4, and 5, where the SNR is set to be 15dB. From this figure, it is seen that the simulated value of SINR loss gradually tends to the derived value in (\ref{gamma}) as the number of antennas at Alice increases. Even in the case of small number of antennas at Alice, the SINR loss difference between simulated and derived is still only about 0.125dB, which is substantially small. This further verifies the validity of the derived expression in (\ref{gamma}).

Fig.~\ref{SR-L-60-120} shows the curves of SR versus number of quantization bits ranging from 1 to 8 for three typical SNRs: 0dB, 15dB, and 30dB, where $N_a = 16$. From this figure, it is clearly seen that there is a certain loss on SR for the small number of quantization bits, i.e., $L=1$ or $2$.  Observing this figure, a 3-quantization-bit phase shifters at Alice will lead to a SR performance loss less than 0.1 bits/s/Hz.

Fig.~\ref{SR-L-Na:SNR} shows the curves of SR versus number of quantization bits  for four different numbers of antennas at Alice $N_a$ : 4, 16, 64, and 256 with three typical SNRs: 0dB, 15dB, and 30dB.
The solid lines represent the SR in the absence of QE, and the dotted lines represent the SR in the presence of QE for different $N_a$.
It can be seen from the figure that three-quantization-bit achieves a SR performance  loss of less than 0.1 bits/s/Hz regardless of the number of transmit antennas at Alice.

In summary, there exists QE in the AB structure due to finite-quantized phase shifters, which will result in a substantial performance loss. In general, from the above simulation results and derived SINR performance loss expression as shown in (\ref{gamma}), we find an important fact that 3, 4, and 5 are sufficient for the number of quantization bits on RF phase shifter such that a performance loss due to QE can be neglected. The derived simple expression in (\ref{gamma}) can be approximately used to assess the SINR performance loss at Bob. Additionally, this expression also holds for even small number of transmit antennas at Alice although it is derived under the condition that the number of antennas at Alice tends to large-scale. This expression can be directly applied in the HAD structure to evaluate the SINR loss.

\section{Conclusion}\label{S5}
In this paper, we have made an investigation of the impact of QE caused by finite-quantized phase shifters of AB structure on performance in DM systems. In the presence of QE, the expression of SINR performance loss has been derived to be inversely proportional to the square of sinc function by making use of the law of large numbers in probability theory. From analysis and simulation, we have found that our proposed expression is approximately close to the corresponding simulated result even when the number of antennas at Alice is small-scale. The SINR performance loss is lower than 0.3dB when the number of quantization bits is larger than or equal to 3. As for SR, we can obtain the same result. In other words, when the number of quantization bits is larger than or equal to 3, the SR difference between NQE and QE is less than 0.1 bits/s/Hz. Additionally, the BER performance  is also shown to be intimately related to the number of quantization bits. A large $L$ means a good BER performance along the desired direction. Otherwise, a small $L$ means a poor BER performance along the desired direction.  Considering the derived SINR performance loss holds for small-scale number of antennas at Alice in AB structure, it is sensible to extend it to a HAD beamforming structure with finite-quantized phase shifters in diverse scenarios for future wireless communications.

\ifCLASSOPTIONcaptionsoff
  \newpage
\fi

\bibliographystyle{IEEEtran}
\bibliography{IEEEfull,cite}

\end{document}